\documentclass{icrc29}
\usepackage{graphicx}
\usepackage{amssymb,amsmath,times}
\usepackage{units}
\usepackage{floatflt}
\usepackage{wrapfig}
\begin{document}
\title[Hadronic multiparticle production in extensive air showers 
and accelerator experiments]{Hadronic multiparticle production in 
extensive air showers and accelerator experiments}
\author[C. Meurer et al.] {C. Meurer$^a$, J. Bl\"umer$^b$, R. Engel$^a$,
        A. Haungs$^a$, M. Roth$^b$ 
        \newauthor
	\\
        (a) Forschungszentrum Karlsruhe GmbH, Postfach 3640, 
            D-76021 Karlsruhe, Germany \\ 
        (b) Universit\"at Karlsruhe (TH), Postfach 6980,
            D-76128 Karlsruhe, Germany
        }

\presenter{Presenter: M. Roth (Markus.Roth@ik.fzk.de), \ 
ger-roth-M-abs1-he21-oral}

\maketitle

\begin{abstract}
 
Using CORSIKA for simulating extensive air showers, we study the
relation between the shower characteristics and features of hadronic
multiparticle production at low energies.  We report about
investigations of typical energies and phase space regions of secondary
particles which are important for muon production in extensive air
showers. Possibilities to measure relevant quantities of hadron
production in existing and planned accelerator experiments are
discussed.

\end{abstract}

\section{Introduction}

One of the most promising approaches to determine the energy spectrum and
composition of the cosmic rays with energies above $10^{15}$\,eV
is the measurement of the number of electrons and muons produced in
extensive air showers (EAS). 
However the results of such a shower analysis are strongly dependent on
the hadronic interaction models used for simulating reference
showers \cite{kascade_holger}.  Therefore it is important to study in
detail the role of hadronic interactions and in particular the energy
and secondary particle phase space regions that are most important for
the observed characteristics of EAS.

The electromagnetic component of a shower is well determined by the
depth of maximum and the energy of the shower. Due to the electromagnetic cascade, having a short radiation length of \unit[$\sim 36$]{g/cm$^2$}, any information on the initial distribution of photons produced in $\pi^0$ decays is lost. 
Therefore the electromagnetic shower component depends on the primary
particle type only through the depth of shower maximum.
In contrast, the muon component is very sensitive to
the characteristics of hadronic interactions.  Once the hadronic shower
particles have reached an energy at which charged pions and kaons decay,
they produce muons which decouple from the shower cascade. The muons
propagate to the detector with small energy loss and deflection and
hence carry information on hadronic interactions in EAS.
Due to the competition between interaction and decay, most of the muons
are decay products of mesons that are produced in low-energy
interactions. Therefore it is not surprising that muons in EAS are
particularly sensitive to hadronic multiparticle production at low energy
\cite{EngelISMD1999}. Recent model studies show that even at
ultra-high shower energies the predictions on the lateral distribution
of shower particles depend strongly on the applied low-energy interaction model
\cite{Drescher2004}.

\section{Muon production in extensive air showers}

Motivated by the measurement conditions of the KASCADE array \cite{kascadeNIM}, we
consider showers with a primary energy of $10^{15}$~eV and apply a muon detection threshold of
\unit[250]{MeV}.  Using a modified version of the simulation package
CORSIKA \cite{CORSIKA} we have simulated two samples of 1500 vertical
and inclined ($60^{\circ}$) proton and 500 iron induced showers.  Below
\unit[80]{GeV} the low-energy hadronic interaction model GHEISHA 2002
\cite{GHEISHA} and above \unit[80]{GeV} the high-energy model QGSJET 01
\cite{QGSJET} are applied.  In the following only vertical proton showers
are discussed. The results are very similar for iron induced showers and also for
zenith angles up to $60^{\circ}$.

\begin{wraptable}[8]{r}{0.35\textwidth}{
\vspace*{-4mm}
\caption{\label{ID} Particle types of mother and grandmother particles 
in a vertical proton induced shower at $10^{15}$eV.}
\vspace*{2mm}
\centering
\begin{tabular}{|c|c|c|}
\hline          & mother  & grandmother \\
\hline
\hline pions    & 89.2\% &  72.3\%  \\
\hline kaons    & 10.5\% &   6.5\%  \\
\hline nucleons & -      &  20.9\%  \\
\hline
\end{tabular}}
\vspace{0.5 cm}
\end{wraptable}

\begin{figure}[h]
\begin{minipage}[t]{0.48\textwidth}
        \centering
        \includegraphics[width=\textwidth,bb=  10 20 511 345,clip]{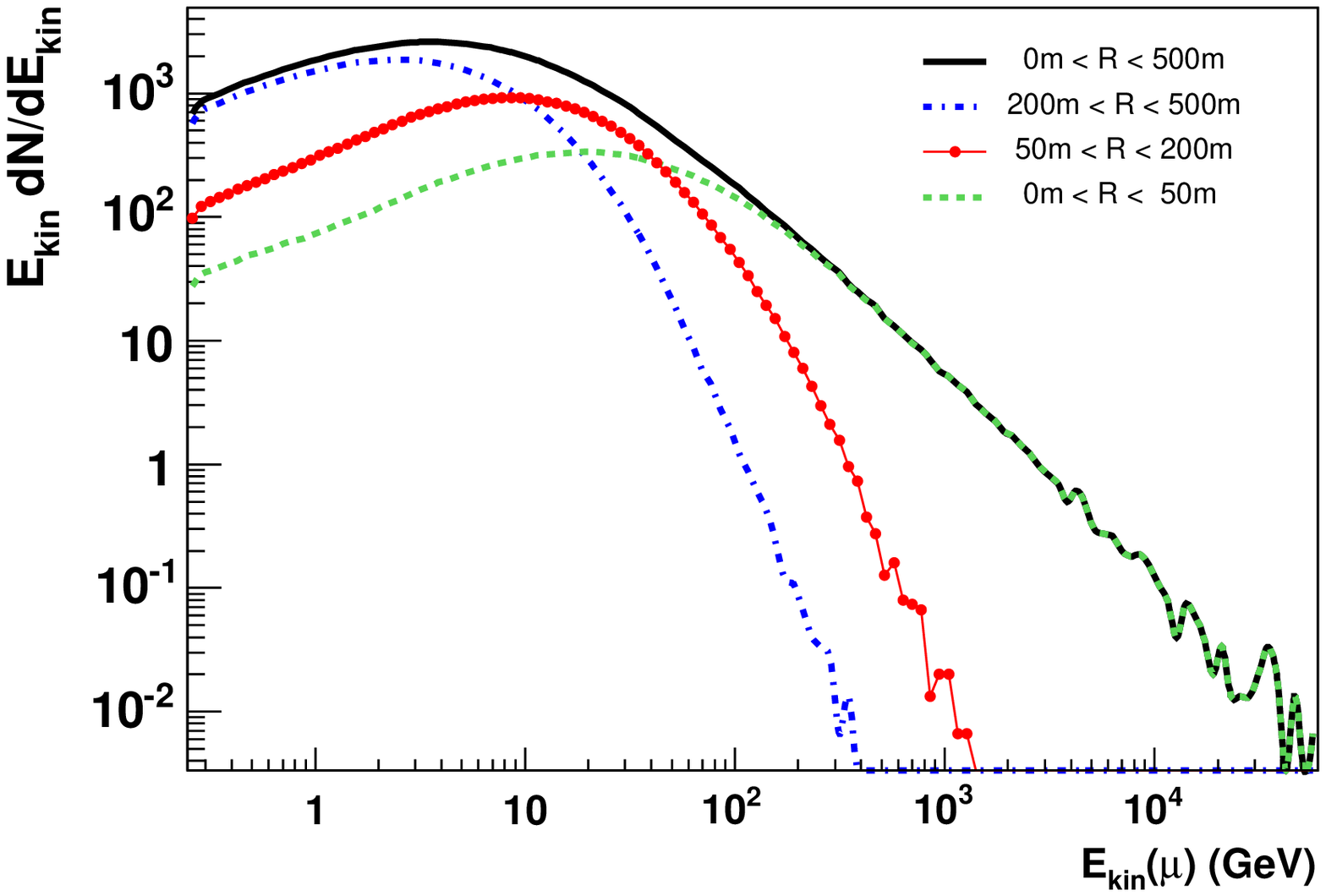}
        \caption{Simulated energy distribution of muons for different lateral distances.}
        \label{Edis_md}
\end{minipage}
\hfill
\begin{minipage}[t]{0.48\textwidth}
        \centering
        \includegraphics[width=\textwidth,bb= 10 20 511 345,clip]{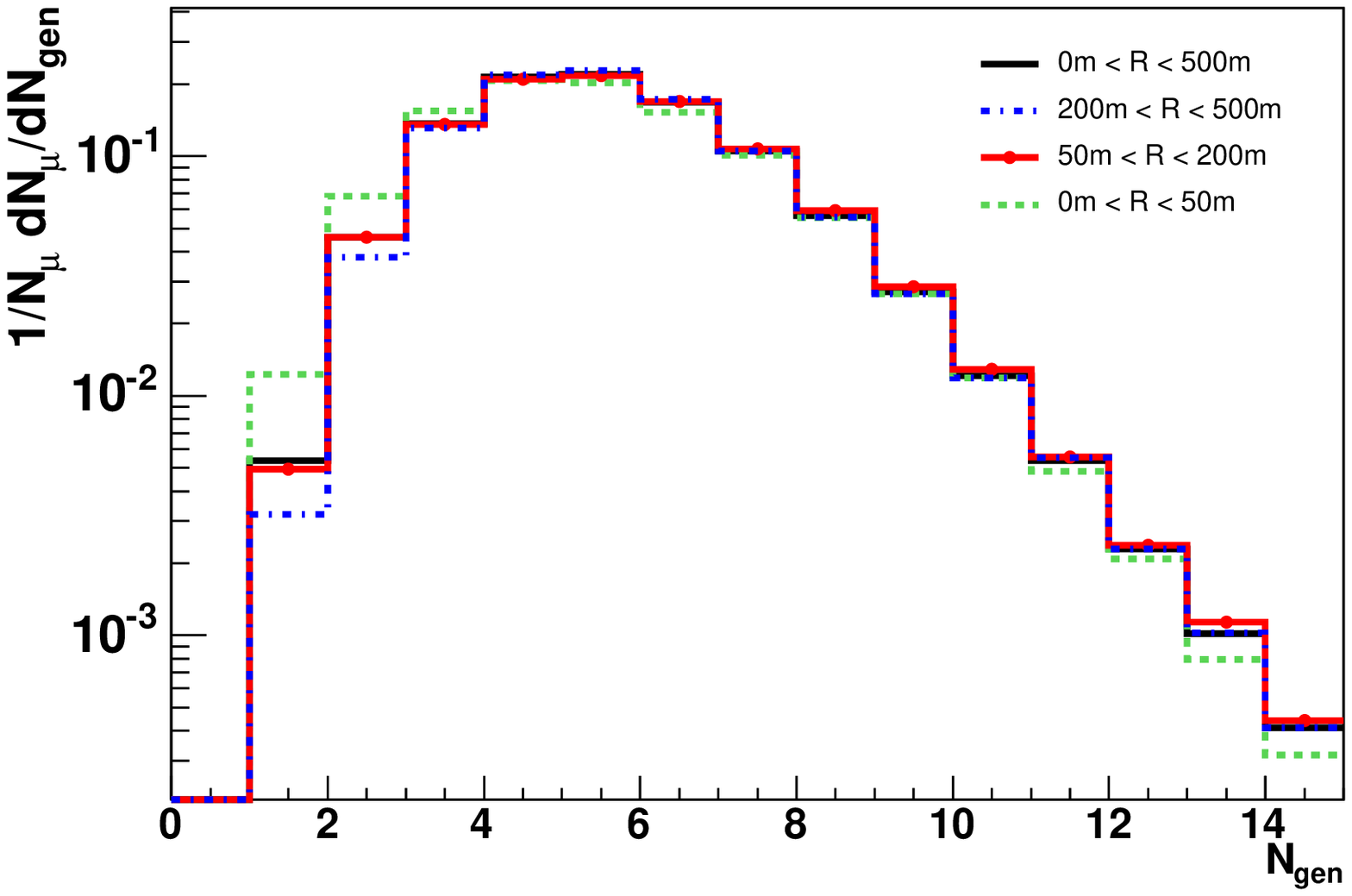}
        \caption{Averaged number of generations before producing a muon visible at ground level (shown for various lateral distances).}
        \label{Ngen}
\end{minipage}
\end{figure}

In Fig.~\ref{Edis_md} the energy distribution of muons at detector level
(\unit[1030]{g/cm$^2$}) is shown for several lateral distance ranges. The
maximum of this distribution shifts to lower energies for larger lateral
distances.  Most likely four to five consecutive hadronic interactions
(number of generations) take place before a hadron decays into a muon,
see Fig.~\ref{Ngen}.  Here and in the following we consider only those
muons that reach the ground level with an energy above the detection
threshold.  The number of generations show no significant dependence on
the lateral distance.  To study the hadronic {\it ancestors} of muons in
EAS, we introduce the terms {\it grandmother} and {\it mother particle}
for each observed muon. The grandmother particle is the hadron inducing
the {\it last} hadronic interaction that finally leads to a meson
(mother particle) which decays into the corresponding muon.  Most of the
grandmother and mother particles are pions, but also about 20\% of the
grandmother particles are nucleons and a few are kaons. Details of the
composition of mother and grandmother particles are given in
Tab.~\ref{ID}.

\section{Energy and phase space regions}

\begin{figure}[h]
\begin{minipage}[t]{0.47\textwidth}
\centering
\includegraphics[width=\textwidth, bb=  10 20 511 345,clip]{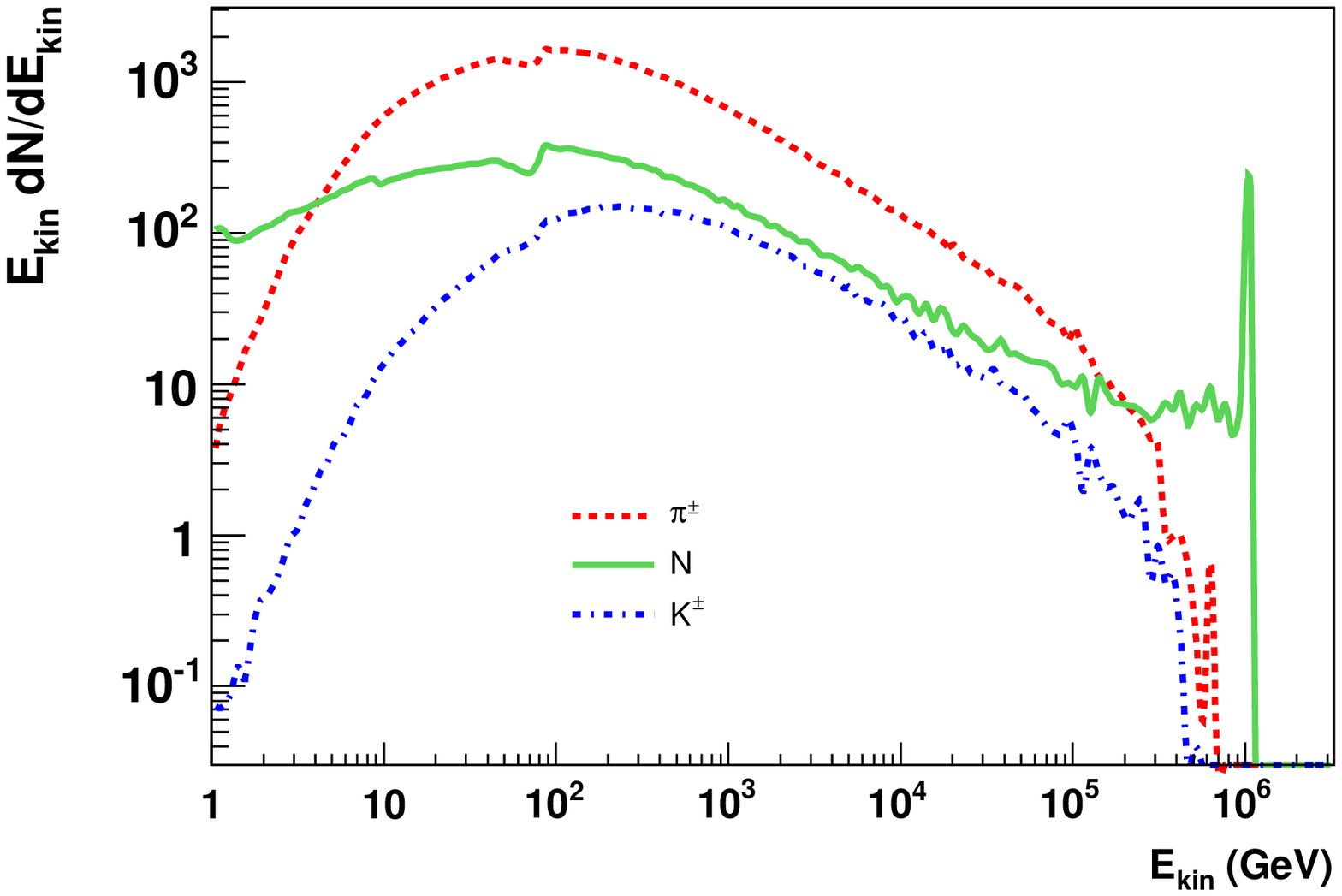}
\end{minipage}
\hfill
\begin{minipage}[t]{0.48\textwidth}
\centering
\includegraphics[width=\textwidth, bb=  10 20 511 345,clip]{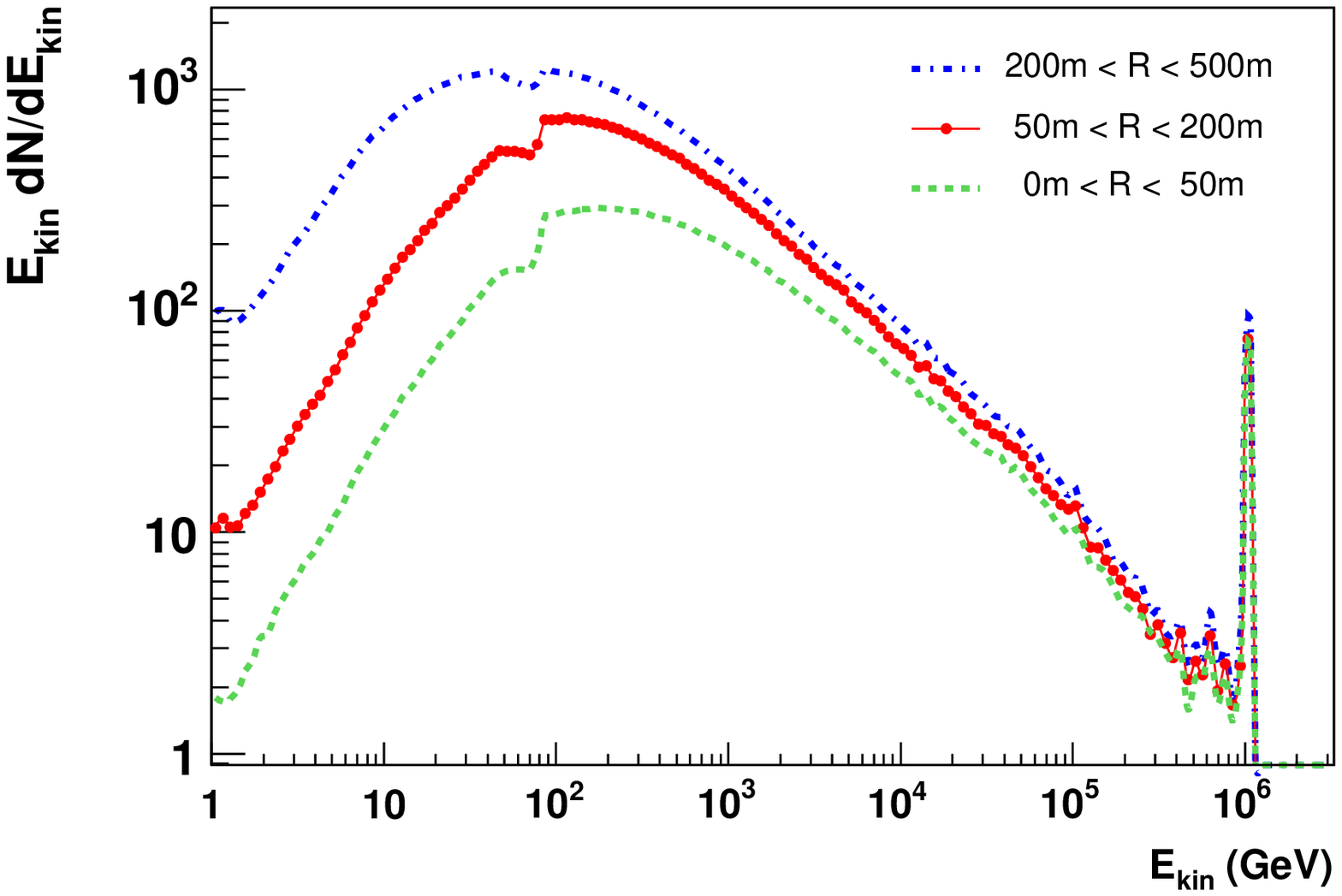} 
\end{minipage}
\caption{\label{Edis_gm} Energy distribution of grandmother particles. 
Left: different grandmother particle types; lateral distance range of muons at ground level: \unit[0-500]{m}. 
Right: different lateral distances; all particle types are summed up. 
}
\end{figure}

The energy spectra of different grandmother particles are shown in
Fig.~\ref{Edis_gm} (left). They cover a large energy range up to the
primary energy with a maximum at about \unit[100]{GeV}. The peak at
\unit[$10^{6}$]{GeV} in the nucleon energy spectrum shows that also a
fraction of muons stems from decays of mesons produced in the first
interaction in a shower.  Furthermore, the step at \unit[80]{GeV}
clearly indicates a mismatch between the predictions of the low-energy
model GHEISHA and the high-energy model QGSJET.  In Fig.~\ref{Edis_gm}
(right) the grandmother particle energy spectrum is shown for different
ranges of lateral muon distance.  The maximum shifts with larger lateral
distance to lower energies. Comparing the {\it last} interaction in EAS
with collisions studied at accelerators, one has to keep in mind that the
grandmother particle corresponds to the beam particle and the mother
particle is equivalent to a secondary particle produced in e.g. a
minimum bias p-N interaction. The most probable energy of the
grandmother particle is within the range of beam energies of fixed
target experiments e.g. at the SPS accelerator at CERN.

\begin{wraptable}[6]{r}{0.54\textwidth}{
\vspace*{-2mm}
\caption{\label{range} For the analysis used energy and lateral distance ranges.
}
\vspace*{2mm}
\centering
\begin{tabular}{|c|c|c|}
\hline energy range             & average energy          & lateral distance range \\
\hline
\hline \unit[80-400]{GeV}       & \unit[160]{GeV}         &  \unit[50-200]{m}  \\
\hline \unit[30-60]{GeV}        & \unit[40]{GeV}          &  \unit[200-600]{m}  \\
\hline
\end{tabular}}
\vspace{0.5 cm}
\end{wraptable}

\newpage

\begin{figure}[ht]
\begin{minipage}[ht]{0.48\textwidth}  
\centering
\includegraphics[width =\textwidth, bb=  10 20 511 345,clip]{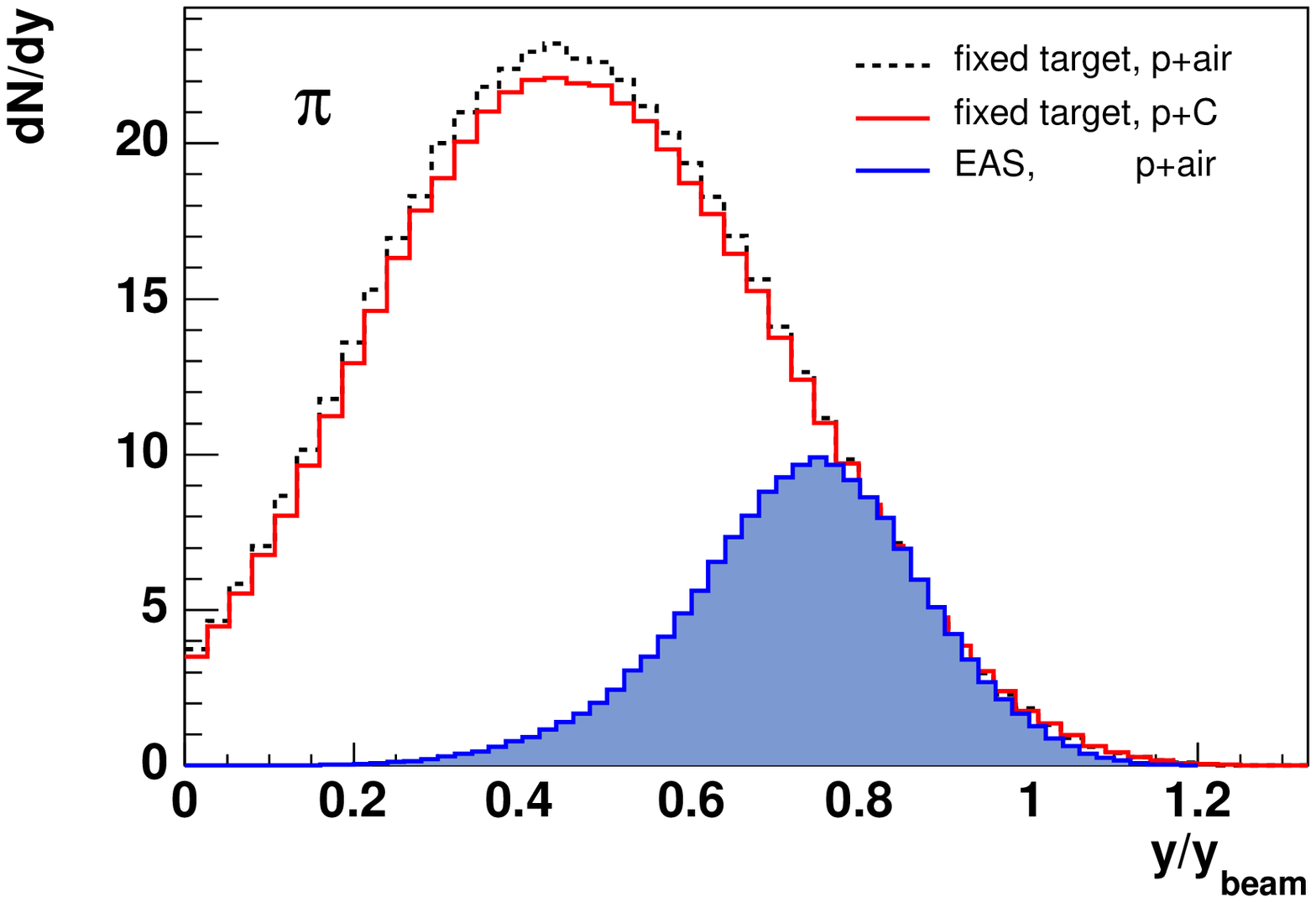}
\end{minipage}
\hfill
\begin{minipage}[ht]{0.48\textwidth}
\centering
\includegraphics[width =\textwidth, bb=  10 20 511 345,clip]{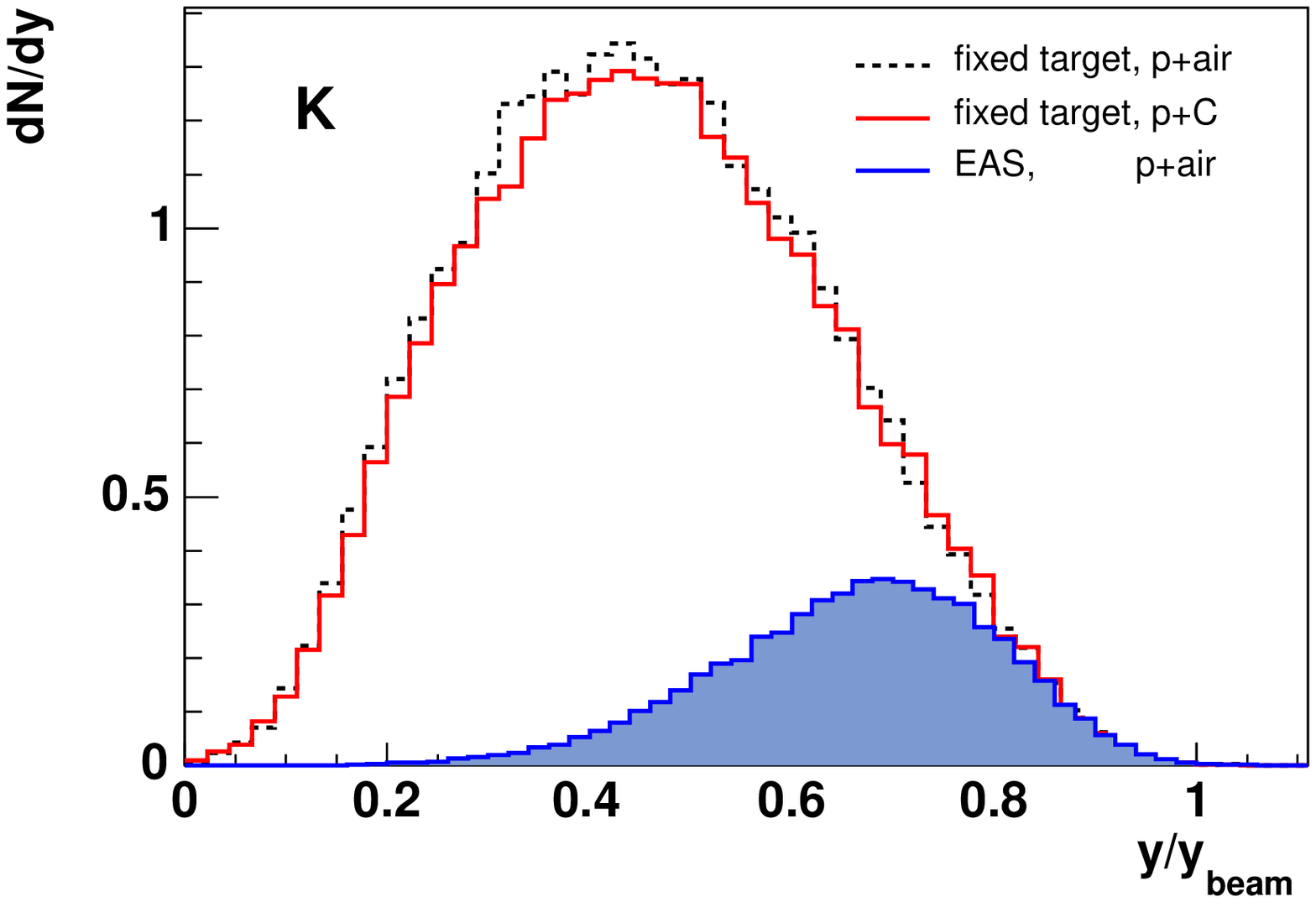}
\end{minipage}

\caption{\label{Rap} 
Rapidity distributions of mother particles (filled
curves) compared with rapidity distributions of secondary particles in
simulated single p+C (solid line) and simulated p+air (dashed line)
collisions. Left: pions. Right: kaons.  The energy range of the
grandmother particle is limited to \unit[80-400]{GeV} and the lateral
distance of the muons to \unit[50-200]{m} to match experimentally
accessible regions. The fixed target collision simulation is done at
\unit[160]{GeV}, corresponding approximately to the mean grandmother
energy.  The rapidity is normalized to the rapidity of the beam and
grandmother particles, respectively.
}
\end{figure}

The further study of the relevant phase space of the mother particles is
done for two different grandmother energy ranges and lateral distance
ranges of muons at ground level, see Tab.~\ref{range}. The lateral
distance ranges are chosen to resemble typical lateral distances
measured at \mbox{KASCADE} and KASCADE-Grande, respectively. 
In Fig.~\ref{Rap} the rapidity spectra of mother particles (left: pions, right: kaons) are
compared to the spectra of secondary particles of proton-carbon
collisions and proton-air collisions simulated with QGSJET labeled as {\it fixed target}. The spectra
of mother particles in air showers are scaled to fit the falling tail of
the fixed-energy collision spectra. No significant differences are found
comparing the rapidity distributions of secondary particles in
proton-carbon and proton-air collisions. As a consequence of the different selection criteria, the forward hemisphere in the mother rapidity spectra is clearly favoured compared to the spectra of secondaries in minimum bias collisions.

\section{Conclusions and outlook}

\begin{wraptable}[6]{r}{0.48\textwidth}{
\vspace*{-8mm}
\caption{\label{ps_range} Phase space regions of hadronic interactions 
relevant for muon production in EAS.}
\vspace*{2mm}
\centering
\begin{tabular}{|c|c|c|}
\hline average energy (GeV) & $y/y_{beam}$ & $p_\perp$ (GeV/c) \\
\hline
\hline 160                  & 0.3 - 1.1    & 0.0 - 0.7  \\
\hline  40                  & 0.3 - 1.1    & 0.0 - 1.0  \\
\hline
\end{tabular}}
\vspace{0.5 cm}
\end{wraptable}

\begin{figure}[h]
\begin{minipage}[t]{0.48\textwidth}
\centering
\includegraphics[width =\textwidth, bb=  10 20 511 345,clip]{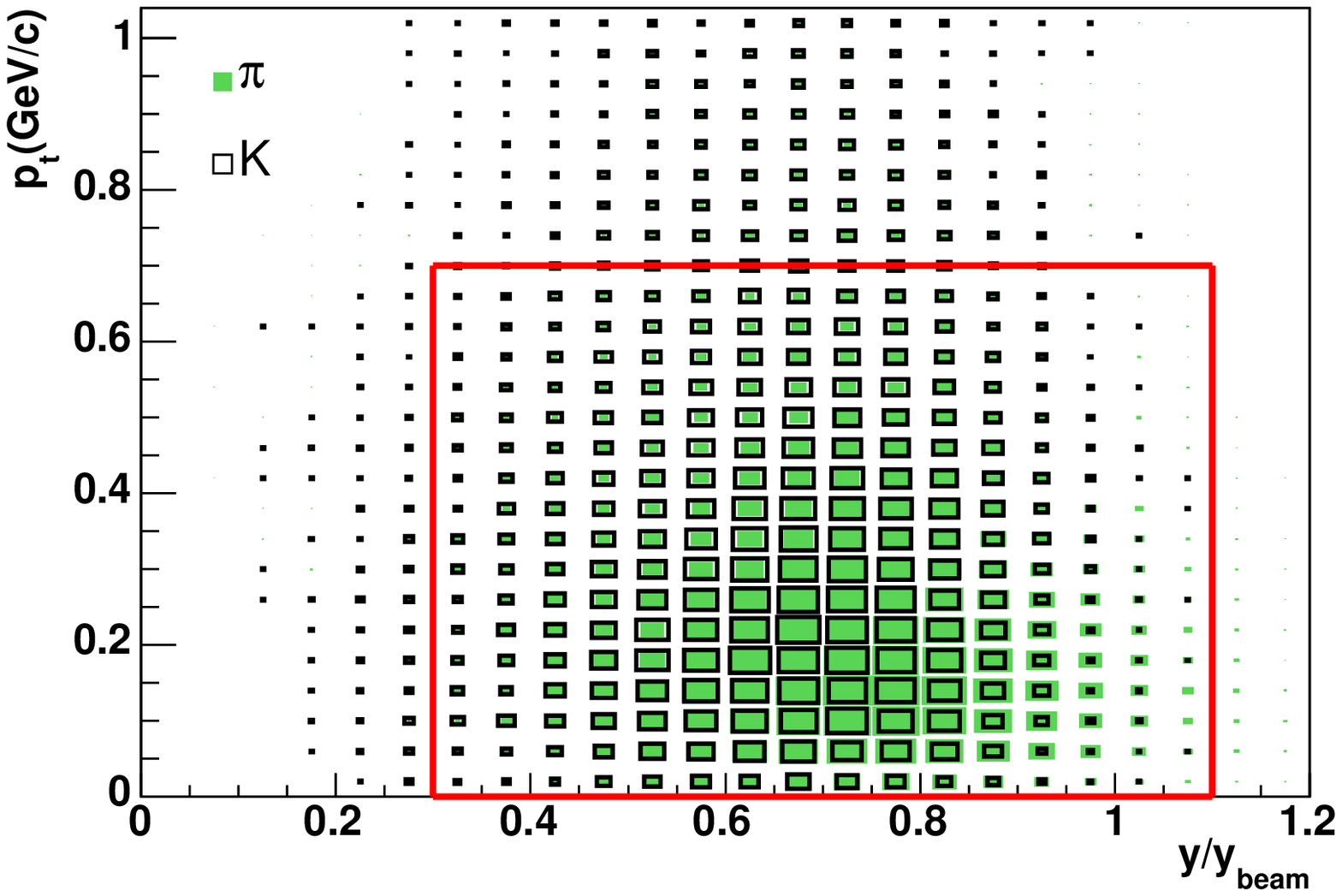}
\end{minipage}
\hfill
\begin{minipage}[t]{0.48\textwidth}
\centering
\includegraphics[width =\textwidth, bb=  10 20 511 345,clip]{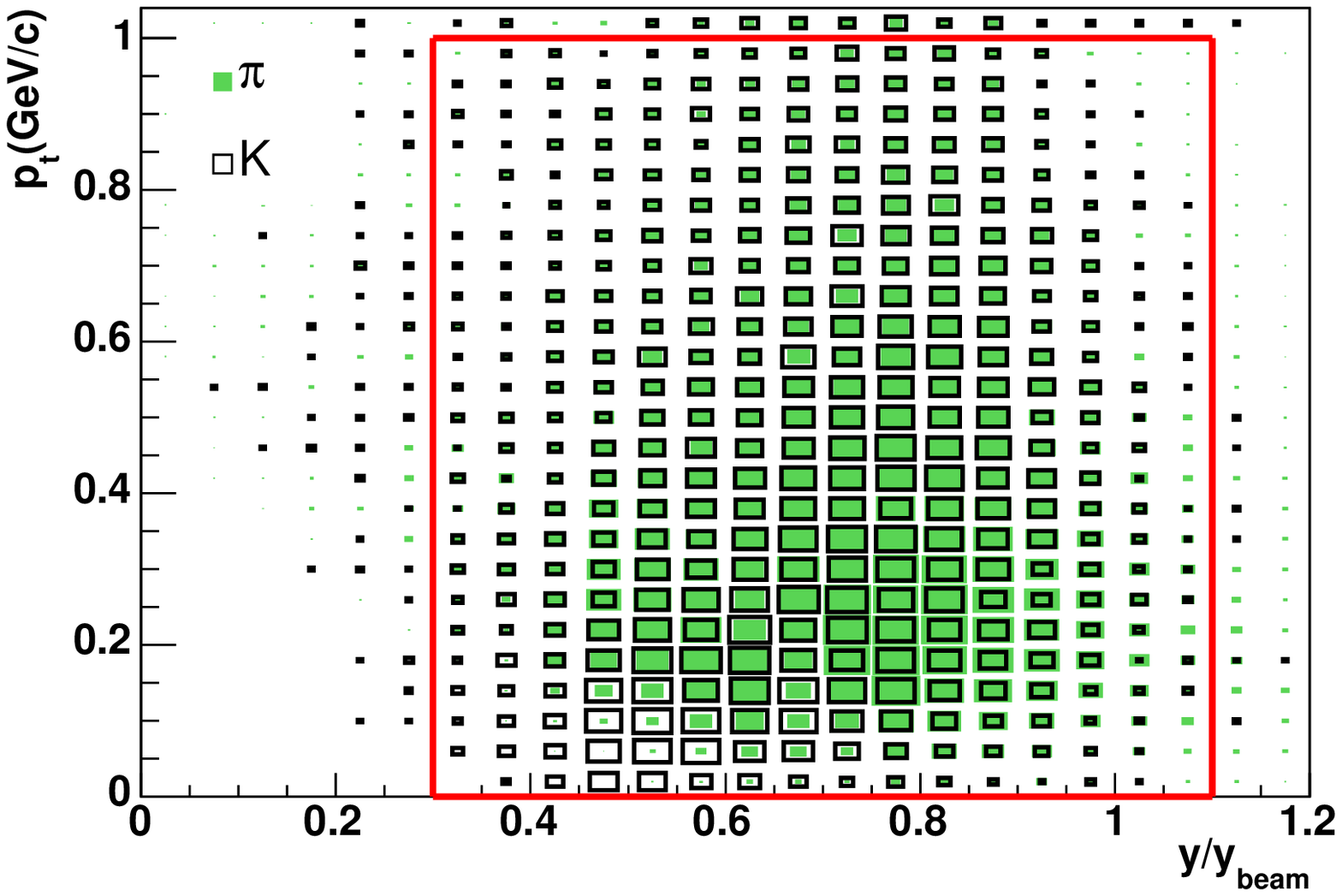}
\end{minipage}
\caption{\label{PS} Phase space of mother particles. Left: grandmother energy range: \unit[80-400]{GeV}. Right: \unit[30-60]{GeV}.
The filled symbols show the distribution for pions, the open symbols for kaons. The large box (red) indicates the most interesting phase space region which includes more than 90\% of this particles.}
\end{figure}

Due to the interplay between decay and interaction of pions and kaons,
low energy hadronic interactions are very important for muon production
in extensive air showers. With increasing lateral distance the mean
energy of these interactions, which are mainly initiated by pions
and nucleons, decreases.  The phase space regions of relevance to EAS
are shown in Fig.~\ref{PS} and summarized in Tab.~\ref{ps_range}.  The
most important interaction energies and phase space regions fall in the
range accessible to fixed target experiments with large acceptance
detectors such as HARP, NA49, and MIPP (see also \cite{Barr2005} and Refs.
therein). Therefore fixed target measurements could be used to improve low energy interaction models that can be independently cross-checked by muon measurements in EAS.

\noindent 
{\bf Acknowledgements:} The authors thank Dieter Heck for 
many fruitful discussions and help with modifying CORSIKA to include 
the muon ancestor information.


\begin{thebibliography}{99}

\bibitem{kascade_holger} T.~Antoni et al. (KASCADE Collab.) 
Astropart. Phys. in press, astro-ph/0505413.
%
\bibitem{EngelISMD1999} 
R.~Engel, T.K.~Gaisser, and T.~Stanev, Proc. of
ISMD, Providence, Rhode Island, August 9-13, 1999, 
World Scientific (2000), p. 457; 
H.J.~Drescher and G.~Farrar, Astropart. Phys. 19 (2003) 235.
%
\bibitem{Drescher2004} H.J.~Drescher, M.~Bleicher, S.~Soff, H.~St\"ocker,
Astropart. Phys. 21, 87-94 (2004);
D.~Heck et al., Proc. of 28th ICRC, Tsukuba, Japan, (2003) p. 279.
%
\bibitem{kascadeNIM} T.~Antoni et al. (KASCADE Collab.) Nucl. Instr. Meth. A 513 (2003) 490.
%
\bibitem{CORSIKA}
D.~Heck, J.~Knapp, J.~Capdevielle, G.~Schatz  and T.~Thouw,
FZKA 6019, Forschungszentrum Karlsruhe, 1998.
%
\bibitem{GHEISHA}
H.~Fesefeldt, Report PITHA-85/02, RWTH Aachen, 1985.
%
\bibitem{QGSJET}
N.N.~Kalmykov, S.~Ostapchenko, and A.I.~Pavlov,
Nucl. Phys. B (Proc. Suppl.) 52B (1997) 17.
%
\bibitem{Barr2005} G.~Barr and R.~Engel, Proc. of 13th 
ISVHECRI, Pylos, Greece, (2004) submitted to Nucl. Phys.~B,
 astro-ph/0504356.
%
\end{thebibliography}
\end{document}